\documentclass[11pt]{article}
\usepackage{authblk}
\usepackage{graphicx}
\usepackage{tabularx} 
\usepackage{booktabs} 
\usepackage{float}

\usepackage[
  left=1.45in,
  right=1.45in
]{geometry}

\usepackage[T1]{fontenc}             
\usepackage{palatino}              

\usepackage[numbers]{natbib}                      

\usepackage{setspace}
\title{Understanding AI Methods for Intrusion Detection and Cryptographic Leakage}

\author{
    \textbf{Reza Zilouchian \quad Micheal Chavez \quad Fernando Koch} \\
    \textsuperscript{1}\textit{Florida Atlantic University, Boca Raton, FL, USA} \\
    \parbox{0.9\textwidth}{\centering
        \texttt{\{rzilouchian2021,mchavez2021,kochf\}@fau.edu}
    }
}

\date{}

\begin{document}

\maketitle
%

\begin{abstract}

We investigate the role of artificial intelligence in cybersecurity by evaluating how machine learning techniques can detect malicious network activity and identify potential information leakage in cryptographic implementations. We conduct a series of experiments using the NSL-KDD and CIC-IDS datasets to evaluate intrusion detection performance across controlled and shifted data environments. Our results demonstrate that AI models can achieve near-perfect detection accuracy within stable network environment. However, their performance declines when exposed to fluctuating or previously unseen traffic patterns. We also observed that learned models identify patterns consistent with side-channel leakage, suggesting that AI can assist in uncovering implementation-level vulnerabilities.

\end{abstract}

%


\section{Introduction}
%

Artificial intelligence (AI) and Machine Learning (ML) models are expected to have a significant impact on cybersecurity through its ability to support cryptographic systems rather than replace their underlying mathematical foundations \cite{MLANDAI,REVIEW}. AI-based techniques have already demonstrated effectiveness in side-channel analysis, where ML models extract patterns from power consumption \cite{powercon}, timing behavior \cite{timing}, or electromagnetic leakage to recover sensitive information \cite{sens}. Recent research further highlights the expanding role of artificial intelligence in cybersecurity applications, including intrusion detection, malware classification, and automated threat analysis \cite{aicyber, zhangAI}.

We outline a series of experiments designed to evaluate the effectiveness and implications of ML techniques within the cybersecurity domain. In particular, we examine whether artificial intelligence can reliably detect malicious network behavior, how robust these models remain when adversarial or mimicry-based feature manipulations are introduced, and whether ML techniques can identify sensitive information leakage through side-channel analysis (SCA). 

If AI systems are capable of interpreting or disrupting encryption implementations, this may reveal important vulnerabilities that could either expose weaknesses in modern encryption schemes or motivate the development of more resilient security systems capable of resisting AI-driven attacks. To explore this possibility, we investigate (i) how effectively ML models can detect malicious network activity?; (ii)  to what extent adversarial or mimicry-based feature manipulations affect the reliability of AI-driven intrusion detection systems?; and; (iii) whether AI techniques can identify sensitive information leakage through side-channel analysis.

This investigation is relevant because understanding how AI behaves under adversarial conditions and whether it can uncover subtle side-channel patterns is therefore essential for evaluating both the defensive and analytical roles of ML in future cybersecurity systems. Modern cybersecurity environments are increasingly dynamic and complex, making traditional rule-based detection systems difficult to maintain and scale. ML approaches offer the potential to automatically identify patterns in large volumes of network and system data, enabling more adaptive and responsive security mechanisms. 

This work contributes to the state of the art by providing:
\begin{itemize}
    \item An evaluation of AI-based intrusion detection models under both controlled and shifted data conditions, highlighting the gap between high training performance and real-world adaptability.
    
    \item An analysis of model vulnerability to adversarial and mimicry-based feature manipulation, demonstrating how targeted perturbations can significantly degrade detection performance.
    
    \item An investigation into the ability of AI systems to identify side-channel leakage patterns in cryptographic operations, showing that learned models can extract partial information that may expose implementation-level vulnerabilities.
\end{itemize}

\label{sec:intro}

\section{Background}
%
%
%

%

Artificial intelligence (AI) has become a critical tool in modern cybersecurity systems due to its ability to process large volumes of data and detect complex patterns that may indicate malicious behavior. As discussed in \cite{aicyber}, AI is widely used to enhance threat detection, including identifying malware, intrusions, and unauthorized access. Prior research demonstrates that AI-based security systems are particularly effective at recognizing anomalies within large datasets, enabling faster and more automated responses than traditional rule-based methods \cite{zhangAI}. These capabilities can reduce human error and improve the overall efficiency of cybersecurity operations.

\begin{figure}[h]
    \centering
    \includegraphics[width=0.6\textwidth]{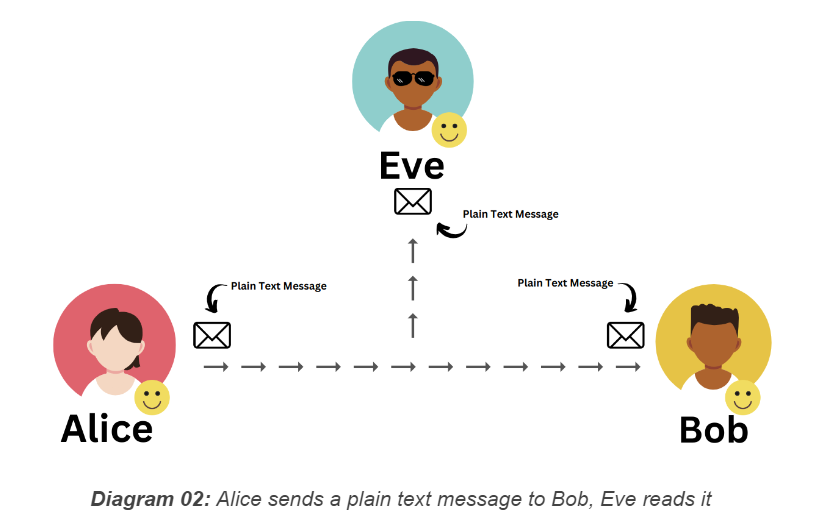}
    \caption{Secure communication between Alice and Bob in the presence of an attacker (Eve).}
    \label{fig:encryption}
\end{figure}

Cybersecurity relies on encryption and decryption techniques to protect sensitive information from unauthorized access \cite{cybersecurityexplaning}. In the modern era of digitalization, vast amounts of data are encrypted to prevent attackers from understanding or exploiting protected systems. Common encryption methods such as the Advanced Encryption Standard (AES) \cite{AES}, Elliptic Curve Cryptography (ECC) \cite{ECC}, and Diffie–Hellman key exchange are widely used in everyday applications to secure communications and data storage. 

Figure \ref{fig:encryption} illustrates this concept, consider two communicating parties, commonly referred to as Alice and Bob, who wish to exchange a message securely. Cryptographic mechanisms such as encryption and hashing protect the transmitted message from an unauthorized third party, often referred to as Eve. Without encryption, attackers could easily intercept sensitive information such as passwords, personal data, or banking records. 

Previous studies such as \cite{SEOAD}, \cite{whatisGAN}, \cite{adNETWORKS}, and \cite{maleware} have explored the role of machine learning in cybersecurity, including approaches where neural networks compete against one another in adversarial settings. These methods have demonstrated effectiveness in intrusion detection, malware classification, and the development of systems capable of identifying or evading detection mechanisms. Such research highlights the growing role of machine learning in both defensive and adversarial cybersecurity applications.

\begin{figure}
    \centering
    \includegraphics[width=0.5\linewidth]{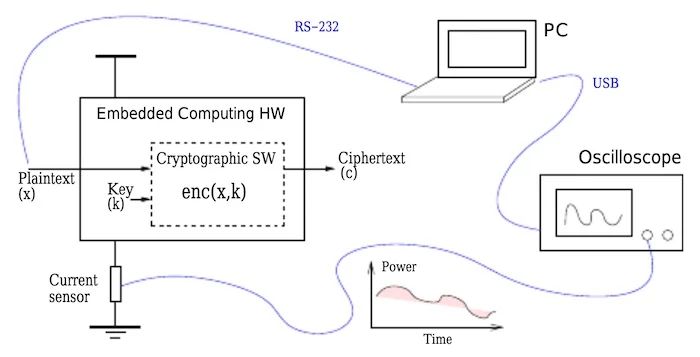}
    \caption{Side Channel Explantion }
    \label{fig:placeholder}
\end{figure}

Figure \ref{fig:placeholder} illustrates the general workflow of side-channel analysis, where power traces are captured from an embedded cryptographic device using measurement equipment such as an oscilloscope and later analyzed to identify correlations between observed signals and internal cryptographic operations \cite{measurement}. During encryption, physical characteristics of the device—such as power consumption, execution timing, or electromagnetic emissions—may unintentionally leak information about intermediate computations \cite{side-channel}. 

Side-channel analysis (SCA) exploits these physical leakages to recover sensitive information without directly breaking the underlying mathematical structure of the encryption algorithm. Instead of attacking the cryptographic algorithm itself, SCA focuses on the implementation of the algorithm within hardware or software systems \cite{sideChannelImple}. One of the most widely studied forms is power analysis, where an attacker records the power consumption of a device while it performs cryptographic operations. Small variations in the measured power traces can reveal information about intermediate computations and secret-dependent operations occurring inside the encryption process \cite{PowerAnalysis}. By applying statistical techniques or machine learning methods to these traces, attackers can identify patterns associated with secret key operations and potentially recover cryptographic key material \cite{MLSIDE}.

\begin{table}[h]
\centering
\caption{Comparison of AI-based cybersecurity research across domains, methodologies, and robustness considerations.}
\label{tab:lit_review}
\small
\resizebox{\linewidth}{!}{
\begin{tabular}{lcccccc}
\hline
Study & Domain & Methodology & Data Type & Robustness Analysis & Side-Channel & Data Shift \\
\hline
\cite{aicyber} 
& Intrusion Detection 
& ML/DL 
& Network Traffic 
& No 
& No 
& No \\

\cite{zhangAI} 
& Malware Detection 
& Deep Learning 
& Binary / Malware Data 
& No 
& No 
& Limited \\

\cite{SEOAD} 
& Adversarial ML (IDS) 
& GAN-based 
& Network Traffic (Synthetic) 
& Yes 
& No 
& No \\

\cite{whatisGAN} 
& Generative Modeling 
& Neural Networks (GAN) 
& Synthetic / General Data 
& Yes 
& No 
& No \\

\cite{adNETWORKS} 
& Network Security (IDS) 
& Deep Learning 
& Network Traffic 
& Partial 
& No 
& Limited \\

\cite{side-channel} 
& Cryptographic SCA 
& Statistical / ML 
& Hardware Traces 
& No 
& Yes 
& No \\

\hline
\textbf{This Work} 
& IDS + SCA-Inspired Analysis 
& ML Models 
& Network Traffic (NSL-KDD, CIC-IDS) 
& \textbf{Yes} 
& \textbf{Yes} 
& \textbf{Yes} \\
\hline
\end{tabular}
}
\end{table}

Table~\ref{tab:lit_review} presents a comparative analysis of representative studies applying artificial intelligence within cybersecurity domains. The table categorizes prior work according to the application domain, machine learning methods used, dataset environments, and whether the approaches consider adversarial robustness, side-channel leakage, or data distribution shifts. As illustrated in the comparison, existing research has extensively explored the use of machine learning for tasks such as intrusion detection, malware classification, adversarial learning, and cryptographic side-channel analysis. However, most studies focus on a single dimension and are typically evaluated under controlled conditions, limiting their ability to reflect real-world system behavior.

This focus creates several limitations. In real-world environments, network traffic patterns are highly dynamic and adversaries may intentionally manipulate features to evade detection systems. While machine learning techniques have demonstrated strong performance in intrusion detection benchmarks, relatively few studies examine how these models behave under adversarial manipulation or shifting data distributions. Similarly, research on side-channel analysis often concentrates on isolated cryptographic experiments rather than evaluating how leakage detection interacts with broader AI-driven cybersecurity systems.

These observations highlight the need for a more integrated analysis of AI within cybersecurity contexts. In particular, it is important to understand how machine learning models perform when faced with changing network environments, targeted feature manipulation, and indirect information leakage through side channels. 
\label{sec:background}

\section{Method}

We introduce an experimentation framework to investigate the role AI within cybersecurity environments. Three experiments are conducted, each aligned with a corresponding research question introduced in Section~\ref{sec:intro}. These experiments aim to evaluate the broader relevance of AI for future security measures and to explore possibile outcomes. The analysis focuses on the findings on collected and synthetic datasets.

\begin{itemize}

\item \textbf{NSL-KDD Network Intrusion Detection Dataset} \cite{NSLKDD}:  consisting of labeled network connection records used to evaluate intrusion detection models. The dataset enables comparative evaluation of classification approaches and provides a structured foundation for assessing detection accuracy and model assumptions.

\item \textbf{CIC-IDS Network Intrusion Detection Dataset} \cite{CICIDS2017}: contains  network traffic data designed for research on intrusion detection, with particular emphasis on packet-level characteristics and payload information. Compared to NSL-KDD, this dataset offers a more detailed representation of network behavior, allowing to evaluate performance differences, and identify potential inefficiencies.

\item \textbf{SCA Dataset}: this synthetic dataset has been created to investigate AI-based side-channel detection, a simulated trace dataset was generated using a controlled leakage model. Each trace represents a time-series measurement of length 400 samples containing additive Gaussian noise. Leakage was introduced at predefined points-of-interest (POIs) by injecting Gaussian-shaped amplitude variations proportional to the Hamming weight. 

\end{itemize}

This controlled simulation framework enables reproducible evaluation of AI models under known leakage conditions while preserving key characteristics of realistic side-channel traces. 

\subsection{Experiment 1: NSL-KDD Intrusion Detection}
\label{subsec:exp1}

This experiment aims to establish NSL-KDD dataset as a baseline for comparison in Experinment 2. This experiment prioritizes $src_bytes$ and $dst_bytes$ because they represent the volume of data sent between source and destination nodes, making them key features associated with network attack behavior. A two-phase learning approach was applied to assess model performance over time. 

Firstly, models were trained on the first half of the dataset, where they learned patterns from labeled data and achieved strong detection performance. Next, these models were tested on the remaining half of the dataset to simulate real-world conditions with unseen data. 

\begin{figure}[H]
    \centering
    \includegraphics[width=0.5\linewidth]{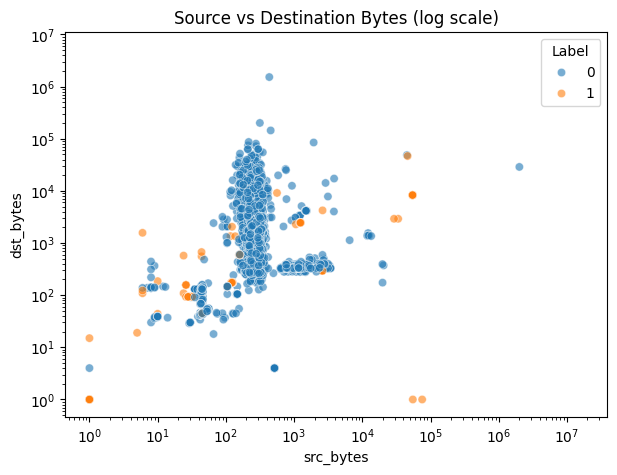}
    \caption{Source vs. destination byte counts (log scale) showing clustered normal traffic and disperse attack activity.}
    \label{fig:scatter}
\end{figure}

Figure \ref{fig:scatter} depicts samples labeled $0$ represent normal network behavior, while samples labeled $1$ correspond to attack activity, with several attack points appearing as outliers demonstrating detectable deviations in traffic patterns.

\begin{figure}[!htbp]
    \centering
    \includegraphics[width=0.8\linewidth]{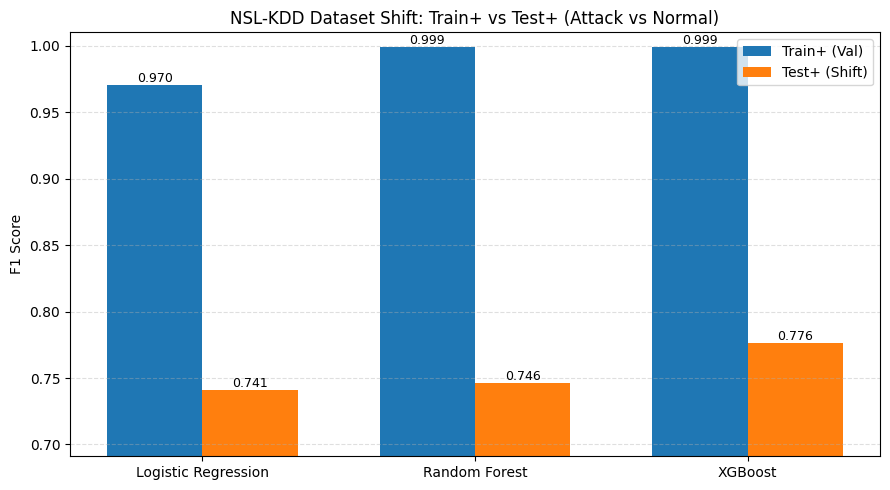}
    \caption{Model performance comparison between training and shifted testing data}
    \label{fig:dataset}
\end{figure}

Figure \ref{fig:dataset} examines whether machine learning models can detect malicious content across different phases, where the training is examining the answers  the shift simulates real-world network conditions. 

\subsection{Experiment 2: Detection Sensitivity Analysis}

This experiment evaluates the sensitivity of the intrusion detection model to direct targeted manipulation of the most influential features. Some of the most influential features consist of $src_bytes$, $serror_rate$, $same_srv_rate$, and $dst_host-level$.

\begin{figure}[H]
    \centering
    \includegraphics[width=0.55\linewidth]{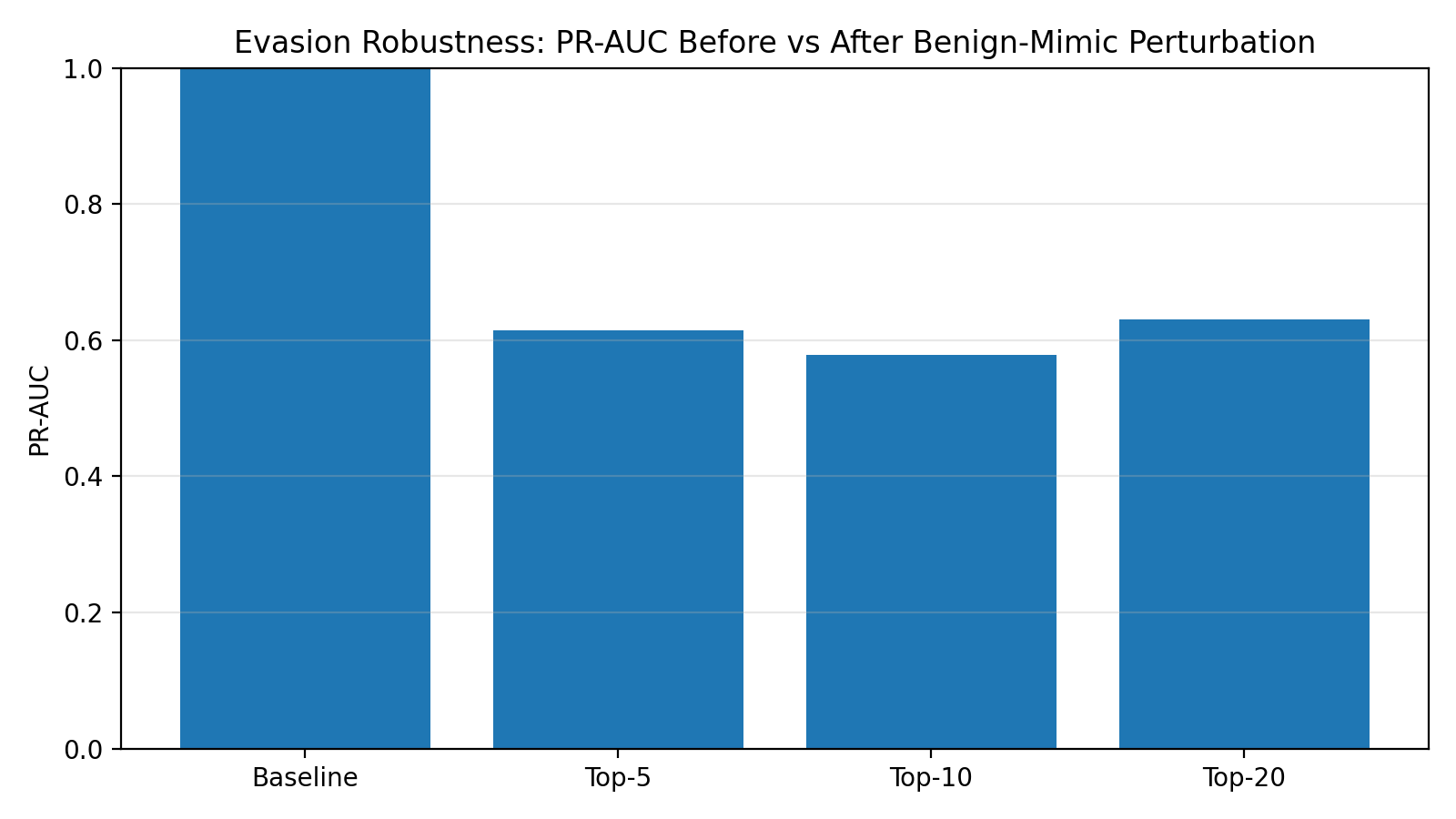}
    \caption{Model performance comparison between training and shifted testing data}
    \label{fig:dataset}
\end{figure}

Figure~\ref{fig:dataset} illustrating the top 5–20 features ranked by importance. The aim of this experiment is to determine whether detection performance declines with the presences of malicious traffic being modified to mimic benign behavior on the dominant predictors identified in Section \ref{subsec:exp1}. Using the permutation algorithm used in Section \ref{subsec:exp1}, the top $k$ most influential features were identified. Those malicious samples in the test were then perturbed by clipping these features to the fifth - 95th percentile observed in benign traffic, simulating mimicry. 

\begin{figure}[H]
    \centering
    \includegraphics[width=0.8\linewidth]{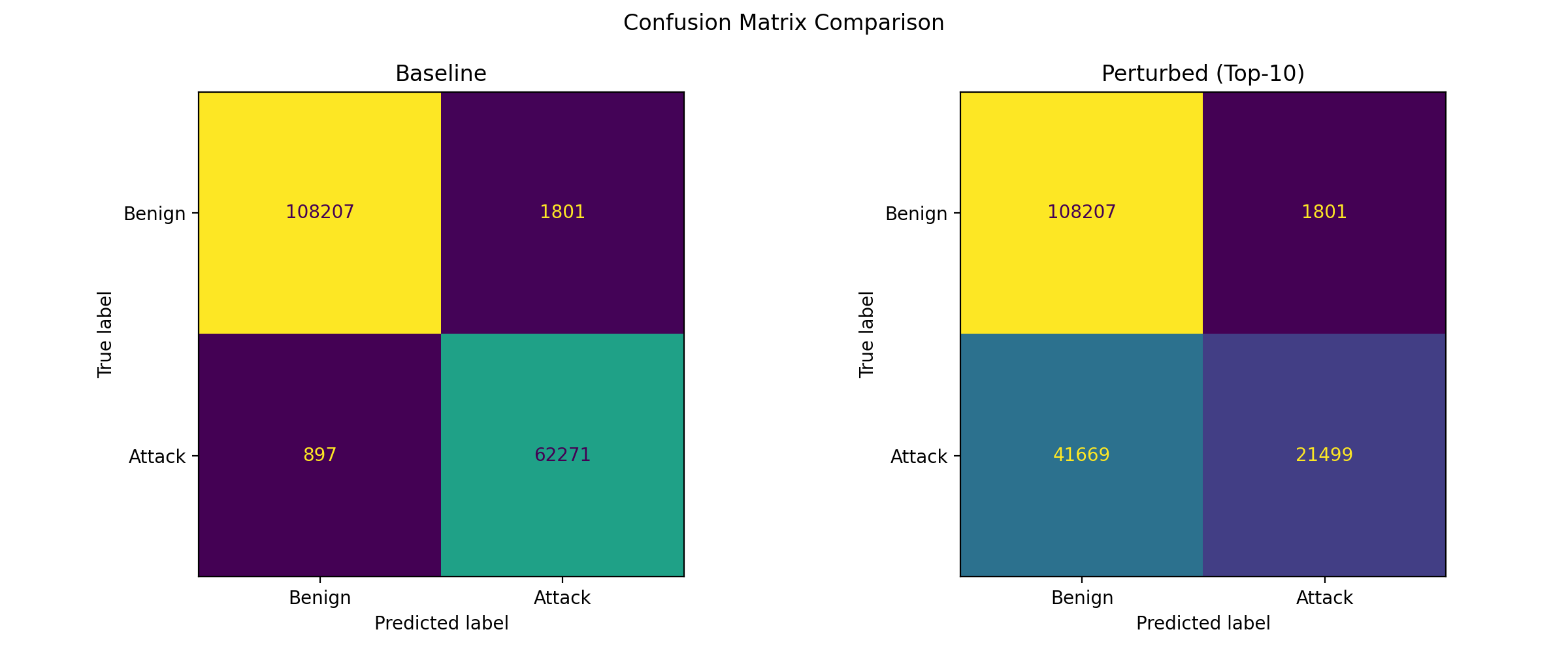}
    \caption{Model performance comparison between training and shifted testing data}
    \label{fig:heat}
\end{figure}

Figure \ref{fig:heat} depicts the evaluation of the trained classifier w on the modified mimicry dataset. Performance was measured using the Precision-Recall Area Under the Curve (PR-AUC) , along with supporting recall and Confusion Matrix Analysis.

\subsection{Experiment 3: Side Channel Analysis}

This experiments aims to assess  ability of AI-based models to identify leakage patterns within a simulated side-channel dataset. The side-channel analysis experiment was performed on a simulated AES-256 encryption implementation, as discussed in Section~\ref{sec:background}, with AI models analyzing leakage patterns produced during the encryption process. Machine learning models were applied to the traces generated during the encryption process to learn data if it can unintentionally reveal information through indirect contact. As shown in 

\begin{figure}
    \centering
    \includegraphics[width=0.90\linewidth]{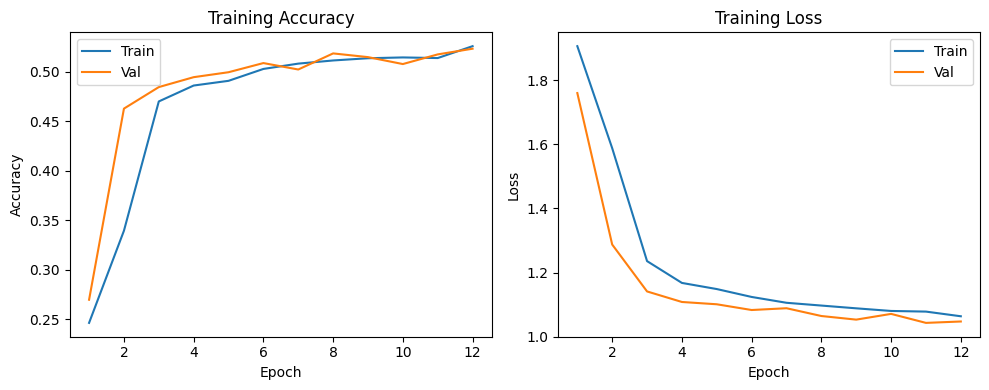}
    \caption{Training and validation accuracy and loss across epochs.}
    \label{fig:training}
\end{figure}

Figure~\ref{fig:training} demonstrates hwo the accuracy slowly improves in epochs, stopping at approximately $50$\%, suggesting that AI models capture partial leakage information.

\begin{figure}
    \centering
    \includegraphics[width=0.45\linewidth]{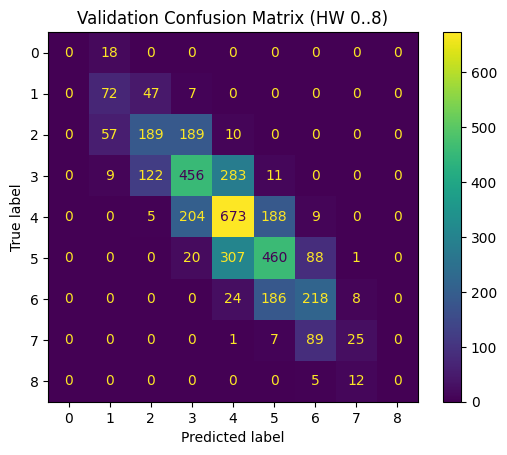}
    \caption{Validation confusion matrix showing classification of leakage classes.}
    \label{fig:confusion}
\end{figure}

Figure \ref{fig:confusion} shows that the models were able to differentiate the leakage classes of the Hamming Weights.

\label{sec:method}

\section{Analysis}
%
%

The results of these experiments highlight both the potential and limitations of AI-driven approaches within cybersecurity systems. The analysis demonstrates that machine learning models can effectively identify malicious network activity when trained within stable and well-defined environments. The findings also reveal that detection performance becomes less reliable when network conditions change or when adversarial manipulation alters key features associated with attack behavior.

Collectively, these findings emphasize that while AI can significantly enhance cybersecurity capabilities, its effectiveness depends on model robustness, data diversity, and careful consideration of adversarial conditions. As AI continues to be integrated into security infrastructures, understanding these strengths and limitations is essential for developing more resilient and adaptive cybersecurity systems.


\subsection{Finding 1: Performance Sensitivity to Data Distribution}

AI-driven intrusion detection systems can achieve high detection accuracy in controlled environments but exhibit reduced reliability when evaluated under shifted or unseen network conditions.

The analysis shows that machine learning models perform strongly when trained and evaluated within stable data distributions. In controlled training conditions, models were able to distinguish malicious traffic from benign activity with near-perfect accuracy, indicating that supervised learning approaches are effective at capturing recognizable attack patterns. Visual inspection of network features such as source and destination byte counts further revealed that malicious traffic often forms distinguishable outliers relative to normal traffic clusters.

However, when the same models were evaluated on previously unseen data representing more realistic network conditions, performance declined significantly. Detection accuracy dropped to approximately $70\%$, suggesting that models trained on a specific dataset may struggle to generalize to new or evolving traffic patterns. This reduction highlights an important limitation of AI-based intrusion detection systems: their reliance on training distributions that may not fully represent the variability of real-world network environments. These results emphasize the importance of dataset diversity and adaptive model training to ensure reliable performance across dynamic cybersecurity contexts.

\subsection{Finding 2: Feature-Level Vulnerabilities in AI-Based Detection}

High-performing intrusion detection models can be highly vulnerable to targeted manipulation of a small set of influential features.

The results demonstrate that intrusion detection models often rely heavily on a limited number of dominant network features when distinguishing between benign and malicious traffic. Under baseline conditions, the classifier achieved a near-perfect Precision–Recall Area Under the Curve (PR-AUC) of $0.9983$, indicating strong separation between attack and normal traffic patterns.

However, when adversarial mimicry was introduced by perturbing the most influential features, detection performance deteriorated substantially. PR-AUC dropped to $0.6148$ when the top five features were manipulated, reached a minimum of $0.5790$ under manipulation of the top ten features (a reduction of $0.4193$ from baseline), and slightly recovered to $0.6307$ when the top twenty features were modified. Correspondingly, the recall for malicious traffic decreased from approximately $99\%$ under baseline conditions to only $34\%$ under the most disruptive perturbation scenario.

This degradation resulted in more than $41{,}000$ missed attacks during evaluation, illustrating how small changes to dominant features can significantly disrupt model decision boundaries. These findings suggest that high predictive performance alone may conceal underlying structural weaknesses in machine learning models, particularly when they rely heavily on a small subset of predictive attributes.

\subsection{Finding 3: AI-Based Identification of Side-Channel Leakage}

Machine learning models can identify leakage patterns in cryptographic traces, indicating their potential for automated side-channel vulnerability discovery.

The analysis of simulated side-channel traces indicates that machine learning models are capable of learning patterns associated with cryptographic leakage. Although classification accuracy stabilized at approximately $50\%$, this performance level still demonstrates that the model was able to extract partial information from noisy time-series traces generated during encryption operations.

Even partial leakage detection can be significant in cryptographic contexts, as it may reduce the computational complexity required to recover secret key material. The confusion matrix further confirms that the model was able to distinguish between leakage classes corresponding to different Hamming weights, a commonly used intermediate representation in side-channel cryptanalysis. Identifying these patterns helps isolate informative points of interest within the trace data, which can assist in recovering secret-dependent information.

It is important to note that the traces used in this study were generated under controlled simulation conditions. Real-world measurements typically contain significantly higher noise levels and additional environmental variability, meaning that model performance may differ when applied to physical cryptographic devices. 
\label{sec:analysis}

\section{Conclusions}
%
%

This work contributes to a deeper understanding of how AI interacts with modern cybersecurity systems by  providing an integrated evaluation of intrusion detection performance, feature-level vulnerabilities, and side-channel leakage analysis, 

We demonstrated both the strengths and limitations of artificial intelligence within modern cybersecurity systems. Across multiple evaluation settings, AI-based models show strong capability in detecting malicious network activity and identifying patterns associated with cryptographic leakage when operating within stable and controlled environments. These results confirm the value of machine learning as a powerful tool for automated intrusion detection and security analysis. However, the findings also reveal important limitations. Model performance declines when network conditions change or when dominant features are intentionally manipulated, indicating that many detection systems rely heavily on a narrow set of predictive attributes. This sensitivity exposes structural vulnerabilities that adversaries could exploit to evade detection mechanisms.

In addition, the side-channel analysis demonstrates that machine learning models are capable of identifying informative leakage patterns within cryptographic traces. Even partial information recovery from noisy signals suggests that AI can assist in uncovering implementation-level vulnerabilities in cryptographic systems. These results highlight the dual role of AI in cybersecurity: while it strengthens detection capabilities, it may also accelerate the discovery of weaknesses in poorly protected implementations.

Overall, the results emphasize that the effectiveness of AI-driven cybersecurity systems depends on robustness, data diversity, and resilience to adversarial manipulation. Developing models that remain reliable under shifting network conditions and adversarial inputs is therefore essential for the safe deployment of AI in security-critical environments.

Future work can extend this framework by incorporating additional datasets and models, as well as by evaluating AI-based side-channel analysis on physical hardware implementations. Such investigations would provide a more realistic assessment of leakage detection under noisy real-world conditions and further clarify the risks and opportunities associated with AI-driven cybersecurity techniques. Together, these directions can support the development of more resilient and adaptive security systems capable of operating reliably in increasingly complex digital environments.
\label{sec:conclusions}


\bibliographystyle{IEEEtran}
\bibliography{references}

\end{document}